\documentclass[conference]{IEEEtran}
\IEEEoverridecommandlockouts
\usepackage{cite}
\usepackage{amsmath,amssymb,amsfonts}
\usepackage[bookmarks=false]{hyperref}
\usepackage{graphicx}
\usepackage{textcomp}
\usepackage{xcolor}
\usepackage{xcolor,colortbl}
\usepackage{graphicx,epstopdf}
\usepackage{subcaption}
\usepackage{balance}  
\usepackage{hyperref}
\usepackage{booktabs} 
\usepackage{amsmath}
\usepackage{dsfont}
\usepackage{algorithm, algpseudocode}
\usepackage{bm}
\usepackage[show]{chato-notes}
\usepackage{multirow}
\usepackage{enumitem}
\usepackage{textcomp}

\newcommand{\dummy}{{\texttt{Dummy}}}
\newcommand{\avg}{{\texttt{AvgSkill}}}
\newcommand{\linear}{{\texttt{Linear}}}
\newcommand{\rf}{{\texttt{RndFrst}}}

\newcommand{\nn}{{\texttt{NN}}}
\newcommand{\logistic}{{\texttt{Logistic}}}
\newcommand{\soft}{{\texttt{NNSoftmax}}}

\newcommand{\linearplus}{{\texttt{Linear$^{+}$}}}
\newcommand{\rfplus}{{\texttt{RndFrst$^{+}$}}}

\newcommand{\nnplus}{{\texttt{NN$^{+}$}}}
\newcommand{\logplus}{{\texttt{Logistic$^{+}$}}}
\newcommand{\softplus}{{\texttt{NNSoftmax$^{+}$}}}
\newcommand{\delal}{{\texttt{Delalleau$^{+}$}}}

\newcommand{\fone}{\textit{F1}}

\newcommand{\spara}[1]{\smallbreak\noindent{\bf{#1}}}

\newcommand{\para}{\smallbreak\noindent}
\newcommand{\etal}{{et al.}}

\newcommand*{\belowrulesepcolor}[1]{%
	\noalign{%
		\kern-\belowrulesep
		\begingroup
		\color{#1}%
		\hrule height\belowrulesep
		\endgroup
	}%
}
\newcommand*{\aboverulesepcolor}[1]{%
	\noalign{%
		\begingroup
		\color{#1}%
		\hrule height\aboverulesep
		\endgroup
		\kern-\aboverulesep
	}%
}

\newcommand{\squishlist}{\begin{list}{$\bullet$}
  { \setlength{\itemsep}{0pt}
     \setlength{\parsep}{3pt}
     \setlength{\topsep}{3pt}
     \setlength{\partopsep}{0pt}
     \setlength{\leftmargin}{1.5em}
     \setlength{\labelwidth}{1em}
     \setlength{\labelsep}{0.5em} } }
\newcommand{\squishend}{
  \end{list}  }

\def\BibTeX{{\rm B\kern-.05em{\sc i\kern-.025em b}\kern-.08em
    T\kern-.1667em\lower.7ex\hbox{E}\kern-.125emX}}
\begin{document}

\title{Competitive Balance in Team Sports Games
}

\author{\IEEEauthorblockN{Sofia Maria Nikolakaki$^\dagger$}
\IEEEauthorblockA{\textit{Computer Science Department} \\
\textit{Boston University, Boston, MA, USA}\\
smnikol@bu.edu\vspace{-.25in}}
\and
\IEEEauthorblockN{Ogheneovo Dibie, Ahmad Beirami, Nicholas Peterson,\\ Navid Aghdaie, Kazi Zaman}
\IEEEauthorblockA{\textit{EA Digital Platform -- Data \& AI} \\
\textit{Electronic Arts, Redwood City, CA, USA}\\
}
\thanks{$^\dagger$The work of SMN was done in part during an internship at EA.}}

\maketitle

\begin{abstract}
Competition is a primary driver of player satisfaction and engagement in multiplayer online games. Traditional matchmaking systems aim at creating matches involving teams of similar aggregated individual skill levels, such as Elo score or TrueSkill. However, team dynamics cannot be solely captured using such linear predictors.
Recently, it has been shown that  nonlinear predictors that target to learn probability of winning as a function of player and team features significantly outperforms these linear skill-based methods.
In this paper, we show that using final score difference provides yet a better prediction metric for competitive balance. We also show that a linear model trained on a carefully selected set of team and individual features achieves almost the performance of the more powerful neural network model while offering two orders of magnitude inference speed improvement. This shows significant promise for implementation in online matchmaking systems.
\end{abstract}


\section{Introduction}
\label{sec:intro}
Video games are now a ubiquitous part of life; it is estimated that there will be over 2.47 billion video gamers worldwide by the end of 2019, according to Statista~\cite{statistica}. 
Furthermore, there has been a big shift toward online gameplay, as the majority of games offer online match capabilities~\cite{online_gameplay}. 
At the forefront of the online experience of video gamers is matchmaking, the process of grouping players or teams into matches. 
It follows a  ``goldilocks'' principle, as research indicates that matches that are neither too difficult nor too easy are key to player stimulation and engagement~\cite{butcher2008pluribus,delalleau2012beyond}. 
Competitive balance is particularly important in team games as it is influenced by both intra- and inter-team dynamics.
A \textit{competitively balanced match}, formally defined in Section \ref{problem_definition},  is one where the distribution of successful scoring events among teams is close \cite{merritt2014scoring}. For example, a soccer match that ends 2-1 is more balanced compared to one that ends 5-0.

The traditional approach for creating competitively balanced matches is to create teams of players with similar aggregate skill ratings. A skill rating is usually a single numeric value derived from a player's prior match history. In team games, a team's skill is an aggregate of the skill ratings of all its players such as the mean/median. Popular skill rating systems today include Elo \cite{elo1978rating}, Glicko \cite{glickman1999parameter}, and Trueskill \cite{herbrich2007trueskill}. 
Although the simplicity of skill-based matchmaking makes it a very attractive choice, multiple studies \cite{delalleau2012beyond,claypool2015surrender} including the research presented in this paper show that this simplicity fails to capture important in-game dynamics that impact match balance. 

In team sports games, teams are comprised of players who play in different roles such as offense, defense, goalkeeper, etc. A single skill rating value indicates a player's overall proficiency and gives no information about expertise in specific roles required for a team. Thus, a team whose players' expertise match the roles they play in will likely dominate an opposing team with little match between player expertise and roles even if both teams have similar skill ratings \cite{wang2015thinking,delalleau2012beyond}. Considering another example, imagine a team game where each team has a player in the forward, midfield and defense role. A team with players whose expertise match the specific role they play in i.e. forward, midfield or defense will likely outperform an opposing team where all players have expertise only in the forward role despite both sets of players having equivalent skill ratings. Therefore, it is not sufficient to use a single skill value to capture the team dynamics in team games.

In this work our contributions are summarized as follows: 
\begin{itemize}
\item We provide a new definition of \textit{competitive balance} for team games, where a match is competitively balanced if the final score difference is concentrated close to zero. Our experiments show that using the proposed definition can lead to ${\scriptsize \sim}15\%$ and ${\scriptsize \sim}2\%$ improved performance over previous definitions in linear models and non-linear models, respectively.

\item We explain and provide insight to a variety of player, team and match features in team sports games.
We design several models to predict competitive balance and demonstrate the definition's utility in a team sports game published by Electronic Arts (EA). 
Our experiments show that using the proposed features can lead up to ${\scriptsize \sim}16\%$ and up to ${\scriptsize \sim}5\%$ prediction performance improvement in linear models and non-linear models, respectively.

\item We demonstrate that using our definition of game balance with the proposed set of features can lead to great computational savings with small predictive performance loss. In particular, the proposed linear model achieves up to ${\scriptsize \sim}100$x computational advantages, particularly at inference times, with less than ${\scriptsize \sim}2\%$ sacrifice in prediction performance compared to non-linear models.
\end{itemize}

\vspace{-0.5pt}

\section{Related Work}
\label{related_work}

\spara{Player, team, and match features for matchmaking:}
While traditional matchmaking systems deal with the creation of 1 player versus 1 player (1v1) matches, the popularity of team games has necessitated the need for systems that can create and match teams. 
These systems are typically team extensions of existing 1v1 skill-based systems, such as Elo \cite{elo1978rating}, Glicko \cite{glickman1999parameter} and Microsoft's TrueSkill \cite{herbrich2007trueskill}. 
For example, a team's skill might be represented by the mean Elo score of all its players. 

A major drawback of these approaches is that they represent a team by a single scalar value\textemdash its skill rating. 
This value, however, does not capture the complex dynamics of competitive team games, such as the distribution of roles in a team,  player play style, team characteristics, etc. 
Recent research has sought to address the drawbacks of considering only skill ratings for team games by modeling team dynamics.
Some researchers \cite{jimenez2011matchmaking,myslak2014developing} have taken advantage of how player skill ratings vary over different roles in a game to create player feature sets comprising role-specific skill levels. 
More recent research \cite{francillette2013players,wang2015thinking} has explored the enrichment of player feature sets with play styles (playing behavior of players during a game). 
For instance, Wang et al. \cite{wang2015thinking} experimentally show on the multiplayer online battle arena (MOBA) game \textit{League of Legends} that teams with a mix of both aggressive and defensive players are more competitive than teams of players of a single style. 

The main issue with these approaches is their focus on creating player-specific features, and using these features to create balanced teams. 
Teams, especially as they become larger, are a lot more complex entities than the sum of their individual members. 
A simple aggregation of the player features does not sufficiently capture match dynamics, such as individual duels between forwards and defenders, or rogue team members, that impact the players' enjoyment. 
Our work focuses on both, creating richer player feature sets, and considering features that capture team dynamics which aren't necessarily tied to a player's characteristics.
 
In this regard, the team profiling approach of Delalleau {\etal} \cite{delalleau2012beyond} is closely related to our work. 
However, in that work the authors define a balanced match to be one where the probability of a team to win is close to 50\%.
Our work extends the work presented by Delalleau {\etal} \cite{delalleau2012beyond} as follows; the authors raise the importance of having richer player feature sets, an issue that we address in this paper by considering generic attributes during the creation of player, team and match-specific features.

\spara{Match balance:}
Balanced matches are a strong indicator of matching opposing teams of similar strength and are more prone to lead to player enjoyment. 
However, a clear and concise definition of match balance is challenging \cite{jaffe2012evaluating}.

The majority of realized matchmaking systems use the average team skill ratings to create matches and assumes that a match is balanced when the opposing teams have close average skill ratings \cite{butcher2008pluribus,leagueoflegends2018}. 
The accuracy of this approach has been challenged by Claypool {\etal} \cite{claypool2015surrender}, who surveyed players participating in skill-based matches and discovered that a majority of these players did not feel that the matches they were involved in were balanced. 
They conclude by stating that player skill ratings are at best useful for ranking players as opposed to creating matches based on them \cite{claypool2015surrender}.

Towards improving the skill rating-based definition of balance, a line of research \cite{chen2016modeling,chen2016predicting,delalleau2012beyond,jaffe2012evaluating} suggests that balance exists when the probability of winning is close to 50\%. 
The recent works of Chen and Joachims \cite{chen2016modeling,chen2016predicting} extend the conventional Bradley-Terry model for the prediction of the winner in $k$v$k$ matches using multi-dimensional representations of the players.
Their experiments show that the proposed model outperforms not only the Bradley-Terry model, but also a variety of baselines.
Delalleau {\etal} \cite{delalleau2012beyond} extend this notion to team games and propose a neural network model to predict the probability of winning. 
Even though such probabilistic models allocate more space and time resources than the simple, but widely used skill-based approach, researchers have began shifting towards this new notion of balance, which is also used for matchmaking in a variety of games \cite{zook2019better}.
Contrary to our setting, all of the aforementioned works approach the generation of matches assessing both probability of winning and player satisfaction.

\vspace{-0.5pt}
\section{Problem Definition \& Notations}
\label{problem_definition}

Throughout the discussion we consider a set of $k$ players $\mathcal{P} = \{P_{j}; j=1,\ldots,k\}$, two opposing teams denoted as $T_{1}$ and $T_{2}$ with team feature sets $t_{1}$ and $t_{2}$ respectively, and a match $M$.

We use real-valued vectors to describe the i) player, ii) team and iii) match feature sets.
In this setting, every player is described by their corresponding feature set, so we use notation $P_{j}$ to represent the player feature set of player $j$.
Furthermore, each team $T_{j\in\{1,2\}}$ is described by its feature set $t_{j\in\{1,2\}}$, respectively.
A match occurring between two opposing teams is described by its feature set $M$, defined by the combination of the feature sets of the opposing teams ($t_{1}$ and $t_{2}$), along with match-specific features ($m$), i.e., $M=(t_{1},t_{2},m)$.

The goal of this paper is to \textit{predict competitive balance} in team online games. 
A match $M$ between teams $T_{1}$ and $T_{2}$ is \textit{competitive balanced}, if the difference between the number of successful scoring events achieved by $T_{1}$ with that of $T_{2}$ approaches zero \cite{csataljay2009performance,gomez2014performance,vaz2011importance}. 
Note that for the general case, a scoring event can be defined broadly and can be a different thing based on the game genre.
For instance, scoring events during a match in FPS and MOBA games can be indicators of the team with more kills, with the most bases captured, or with the most men standing.

\vspace{-0.5pt}
\section{Method}
\label{sec:method}

Our model architecture used for predicting match competitive balance comprises three main phases; (i) the extraction of player features, (ii) the aggregation of these features to form team and match level features, (iii) and the predictor for a balance match.
Note that the presented architecture is a sophisticated extension of the one presented in \cite{delalleau2012beyond}. 
However, our main contribution is in motivating and optimizing the correct metric for competitive balance, and in designing a model with high running time performance using an appropriate set of features, rather than designing a powerful model architecture.

\subsection{Data}
The analysis presented in this paper is based on  data from two team game modes, namely the 3 players versus 3 players (3v3) and the 6 players versus 6 players (6v6) game modes, of an online team sports game published by Electronic Arts, Inc. (EA). 
Both datasets comprise more than 100,000 players and more than 500,000 games, and the balanced samples are generally in similar order to unbalanced samples.
All models are trained and evaluated on subsets of the same 3-month data period.

\subsection{Feature construction}
\label{sec:features}
The conversion of raw attributes to meaningful features used for the prediction of competitive balanced matches occurs during the feature construction phase.

\spara{Player features:}
The set of features that corresponds to player $j$, i.e., $P_j$ is constructed from a player's in-game attributes as described in the game logs. These features are classified into four broad categories as shown in Table \ref{tbl:playerprofile}; i) match experience, ii) role experience, iii) play style, iv) dropout history. 

The match experience category captures general player participation and influence in matches, such as the number of matches a player has played, and the fraction of matches the player has won.
The role experience category captures player experience in specific roles. 
Roles can be either explicitly defined by the game, e.g., forward, defense, etc. or they can be inferred from a player's play style, e.g. a player who saves many scoring attempts may be categorized as an ``defender''. 
Regardless of the role type, maintaining the frequency of a player's involvement with a specific role is a strong indicator of the player's play style. As discussed in Section \ref{related_work}, this insight can be helpful in the creation of balanced teams/matches. 
The play style category covers all actions performed by a player. It differs from the role experience class in that we record statistics about different actions instead of roles. 
Actions generally reflect all micro-level in-game events during a match. 
These include scoring attempts, giveaways, hits, takeaways, etc.
Finally, we consider the dropout history category. 
An essential aspect that leads to competitive balanced matches is ensuring \textit{a priori} that the number of players in each team will remain close throughout the whole match. 
Matches invariably end up unbalanced when players from a particular team quit early.

\begin{table*}[tbp!]
	\centering
	\footnotesize
	\begin{tabular}{lll}
		\hline
		\textbf{$P_{j}$ Feature} & \textbf{Category} & \textbf{Description} \\
		\hline
		num\_matches & Match Experience & Number of matches a player has participated in.\\
		num\_wins & Match Experience & Number of wins a player has had.\\
	    freq\_wins & Match Experience & Ratio of wins to the number of total matches \\
	    & & of a player has participated in.\\ 
		\hline
		num\_role\_i & Role Experience & Number of times a player has played a specific role $i$. \\
		freq\_role\_i & Role Experience & Ratio of times that a player played a specific role $i$ \\
		\hline
		num\_action\_$i$ & Play style & Number of times a player has performed a specific action $i$.\\ 
		avg\_num\_action\_$i$ & Play style & On average, how many times a player performs \\
		& & a specific action $i$ in a match. \\
		\hline
		num\_dropout & Dropout History & Number of times a player has dropped out from a game.\\
		freq\_dropout & Dropout History & Ratio of times a player dropped out from a game.\\
		\bottomrule
	\end{tabular}	
	\caption{A summary of a player feature set $P_{j}$.
		\label{tbl:playerprofile}}
\end{table*}

The player features for all categories described are cumulative and updated in an online fashion as matches are completed. 
This renders player feature sets time-dependent.
It also enables our model to account for recent player activity in making predictions on match competitive balance.

\spara{Team features:}
Team features are based on aggregating individual player statistics.
Given a team, we compute the average value and the standard deviation of its players for each of the player features described in Table \ref{tbl:playerprofile} to create the corresponding team features.

\spara{Match features:}
A match is a complex entity whose performance is determined by both inter- and intra-teams dynamics. 
A match feature set that is only based on aggregating individual player statistics lacks insight on significant aspects of a match that impact competitive balance such as team properties and duels between members of opposing teams.   
Here, the match feature set contains the team features of the opposing teams with additional match-specific features.

First, we use the team features of the opposing teams to create the following feature categories: i) the  absolute difference of each of the average player feature values of the two teams (non-negative value), ii) the difference of the average player feature values of the two teams (can be negative).
The first category aims to capture potential superiority of one team over the other that could lead to an unbalanced match. 
For instance, a team comprising players that have scored many goals in past matches could dominate a team with players that have been less successful at it.
Such phenomena are captured using the difference of the team members' average feature values.
Now even though the absolute value itself demonstrates the existence of a superior team, it does not tell which team that is, which requires the second category of features (signed difference).
These categories provide us with insight into the similarity and differences in team ability. 
For example, a match with a large difference in the average attempts of goals between the two teams will probably be more unbalanced.

Furthermore, we consider features that are based on player information that is available at the time of matchmaking. 
These include the skill ratings of players, their allocated roles in the game and the team sizes. 
We compute various transformations of these features to capture insights on the match composition.
First, as is done in skill-based matchmaking systems, we compute the skill rating of each team by averaging the skill ratings of its players. 
We then compute the difference and absolute skill difference between the two teams.
Other features we compute include the skill difference of the players with the highest skill ratings in each team, the skill difference of the players with the lowest skill ratings in each team as well as the standard deviation of the skill ratings in each team. 
These features showcase if there is a stronger or weaker link within any team that could affect the overall performance, or if one team has a much larger range of skill ratings compared to the other team.

Finally, due to the importance of creating matches fast, many matchmaking systems create teams of different sizes.
This leads to the last set of match features, namely the number of human players in each team. 
In particular, games usually position game bots (short for robots) in roles where human players are missing.
However, the extent to which these can mimic how a human would play the game is rather limited, and they either result in dominating the game, or significantly under-performing.
Therefore, having opposing teams with a similar number of human players is an important indicator of match balance. 

We stress that the described set of features implicitly models synergy among players with different role preferences and playstyles because the corresponding features are the input to a neural network-based model that captures the non-linear dependencies between player features.
We do not present all of the match features due to restricted space; however, in Section \ref{sec:exp}, we do provide the most significant features impacting competitive balance.
\spara{Feature pre-processing:}
Finally, all features are standardized following z-score normalization \cite{zill2011advanced}. 
In total, we propose approximately 100 features for the 3v3 and 6v6 game modes, respectively. 
This number depends on the available set of roles and actions in a game.

\subsection{Prediction model}
\label{sec:predmodel}
This section presents two categories of predictions models used for determining balanced matches.
\spara{Probability of winning prediction model (Delalleau et al.~\cite{delalleau2012beyond}):}
The probability of winning prediction model, proposed by Delalleau {\em et al.}~\cite{delalleau2012beyond}, 
trains a soft classifier that predicts the probability that either team wins given player, team, and match features.
In this case, a match is considered to be balanced if the probability of winning is about $0.5$ for either team.

\spara{Competitive balance prediction model (this work):}
This work proposes using the definition of competitive balance to determine balanced matches.
In particular, the predictor is tasked with regressing the final score difference between the two teams on the match features. The smaller the difference, the more competitive balanced the match is. This score difference value is used to determine match balance via a threshold function.
We motivate the effectiveness of the aforementioned definition in Section ~\ref{sec:esd}, and compare the results with the probability of winning prediction model.
We refer to this model as {\nn} and its succession is as follows:
\para{1)}
The player feature sets $P_{j}$ containing the latest features of each player are retrieved from the database.  
\para{2)}
For each team $T_{j\in\{1,2\}}$ the corresponding team features $t_{j\in\{1,2\}}$ are created.
\para{3)}
The match feature set $M$ combines into a single vector the individual features of each opposing team, along with the additional match-specific features:
$
    M = (t_{1},t_{2},m).
$
\para{4)}
Match features are summarized by a predictor. 
We compare a linear and a two-layer neural network predictor similar to the one in \cite{delalleau2012beyond} with fully connected layers followed by the Rectified Linear Unit (ReLU) activation.

\para{5)}
The output layer of the previous step returns a single, continuous value $r$ representing the final score difference prediction.
Depending on the application, it might be useful to convert the real value into a binary label, denoting whether the match will be balanced or not.
For this purpose, we use function $f:\mathbb{R}\to \{0,1\}$ defined by,
$
    f(r)= \mathbb{I}_{|r|< \theta}(r)
$, 
where $\mathbb{I}$ denotes the indicator function, $r$ is the signed score difference, $\theta$ is a threshold hyperparameter for measuring competitive balance, and where 0 and 1 represent an unbalanced and balanced match, respectively.

\spara{Selecting threshold hyperparameter $\theta$:} In some team sports games if one party leaves the match before it ends then they forfeit. In sports the forfeiting team loses with a predefined score difference, e.g., for soccer and hockey the match ends with 3-0, basketball 25-0, e.t.c. This means that in sports $\theta$ can be clearly defined as the score difference after a forfeit (a forfeited match can be considered unbalanced). Alternatively, $\theta$ could be treated as a hyperparameter that could be tuned for optimizing player engagement and retention.
\spara{Training and validating the prediction models:} The time sensitive nature of our data make them unnameable to a traditional data shuffling and K-fold cross validation procedure. 
Furthermore, recent matches are stronger predictors of competitive balance in upcoming matches than older matches are. 
Therefore, we perform training, validation, and testing as follows. 
Assume a total of matches that occurred within $K$ days that are used for the model's evaluation and hyperparameter tuning.
We use the first $K-3$ days for training, the matches that occurred during days $K-2$ and $K-1$ for validation and the matches of day $K$ for testing.
These base sets allow us to design and evaluate our model in an offline way, before deploying it into the matchmaking system.
Now, we assume that there is a stream of incoming matches arriving at day $K+1$.
We shift the first $K$ days by one, such that the training set contains data from the first $K-2$ days, the validation set includes the next $K-1$ and $K$ days, and the most recent chunk $K+1$ is used for testing.
This procedure continues, and allows the system to continuously update the model using a larger training set, and selecting the most recent validation and testing parameters that capture current tendencies.

\subsection{Competitively balance-based matchmaking}
\label{sec:integration}
In this section, we presented an architecture for predicting competitive balance  to improve matchmaking. 
We emphasize that the results presented in this paper are based on existing player data, but none of the approaches have been deployed to a live matchmaking service.
That said, here we describe how the proposed model can be deployed to a live matchmaking system.
We assume a matchmaking system similar to the ones presented by Delalleau {\etal} \cite{delalleau2012beyond} and Zook et al. \cite{zook2019better}.
Briefly, players enter a queue and the matchmaking system assembles teams using a sampling strategy, calculates the match quality, and either reassembles the teams if the quality is low, or launches the match.
In a similar matchmaking system, the prediction model is integrated with the match quality computation step, and is used as an additional quality assessment addressing the competitive balance of a match.
Predicting whether a match is going to be balanced has low computational overhead, while the prediction model itself can be trained offline.

\vspace{-0.5pt}
\section{Experiments}
\label{sec:exp}
The purpose of this section is to explore the efficiency of our model on real datasets.
Specifically, i) we evaluate and compare the performance of our prediction models to a variety of baseline models, ii) we demonstrate that the definition of competitive balance as a regression problem leads to significant prediction performance improvements, iii) we showcase that using the proposed definition of balance in combination with the proposed features can lead to substantial computational savings, iv) we discuss which features have the most influence on competitive balance in team sports games.   

For context on execution times, our experiments were conducted using single process implementations on a 64-bit MacBook Pro with an Intel Core i7 CPU at 2.6GHz and 16 GB RAM.
All presented models are implemented in Python, using the the scikit-learn \cite{pedregosa2011scikit} and Keras \cite{chollet2015keras} libraries.

\subsection{Baseline methods}
\label{sec:baseline}
We compare the performance of the model {\nn} presented in Section \ref{sec:method} to a variety of baseline methods. 
\spara{{\dummy}:}
{\dummy} is the most naive approach that we consider.
It always predicts the mean of the training set.
{\dummy} corresponds to a competitive balance prediction model.

\spara{{\avg}:}
In the {\avg} approach the match feature set includes only two features , i.e., the two averages of the skill ratings of the players in each team.
These features are used as the input to a linear regression model that predicts the final score difference. 
Note that this baseline corresponds to the currently used single-valued skill aggregation model and corresponds to a competitive balance prediction model.

\spara{{\linear}:}
{\linear} is a linear regression model, where the input match instances comprise all the features presented in Table \ref{tbl:playerprofile}.  
In addition to being a fundamental regression model that is known for its simplicity, linear regression provides insight into the model covariates that explain the variance in the response variable (final score difference). It provides us insights into which explanatory variables are significant in the match balance prediction task.
{\linear} corresponds to a competitive balance prediction model.

\spara{{\rf}:}
Random Forests construct a multitude of decision trees at training time and output the mean prediction of all trees. 
{\rf} corresponds to a competitive balance prediction model.

\spara{{\logistic}:}
This is a logistic regression model that uses a logistic function to model a binary dependent variable.
In particular, it models the probability of a certain class.
{\logistic} corresponds to a probability of winning prediction model.

\spara{{\soft}:}
{\soft} is a model with the same neural network architecture as {\nn} with a single difference.
We replace the final layer with a softmax layer to assign a probability to whether a match is balanced or not. 
{\soft} corresponds to a probability of winning prediction model.

For each of the baseline methods we select the best feature subset using the \emph{recursive feature elimination} method and a statistical feature analysis.
Further details on significant features are provided in Section \ref{sec:sigfeatures}.
All models with the sign $^+$ in their name use their corresponding best subset of features.
Note that we did not perform best subset feature selection for {\dummy} and {\avg} since the features of these models are determined by their definitions.

\subsection{Model Characteristics}
\begin{table}
\centering
\footnotesize
  \begin{tabular}{|l|l|l|l|l|l|l|l|l|l|l|}
    \hline
      \multicolumn{1}{|c|}{\textbf{Model}} &
      \multicolumn{2}{c|}{\textbf{F1}}\\
    & 3v3 & 6v6\\
    \hline
    {\dummy}
    & $0.00\;(\pm 0.00)$ & $0.61\;(\pm 0.01)$\\
    
    {\avg}
    & $0.00\;(\pm 0.00)$ & $0.61\;(\pm 0.01)$\\
    
    {\linear}
    & $0.59\;(\pm 0.26)$ & $0.70\;(\pm 0.08)$\\
    
    {\rf}
    & $0.57\;(\pm 0.17)$ & $0.64\;(\pm 0.07)$ \\
    
    {\nn}
    & $0.62\;(\pm 0.13)$ & $0.73\;(\pm 0.08)$\\
    \hline

    {\linearplus}
    & $0.60\;(\pm 0.27)$ & $0.73\;(\pm 0.12)$\\
    
    {\rfplus}
    & $0.58\;(\pm 0.19)$ & $0.65\;(\pm 0.12)$ \\
    
    {\nnplus}
    & $\mathbf{0.64\;(\pm 0.10)}$ & $\mathbf{0.74\;(\pm 0.09)}$\\
    \hline
  \end{tabular}
  \caption{Training-set performance of models when predicting competitive balance.  
		\label{tbl:trainingresults}}
\end{table}

\label{sec:modchar}
To demonstrate the performances of our prediction models, we use the \emph{F1} metric \cite{powers2011evaluation}.
To support our claims we showcase the results of the training model evaluation along with the corresponding standard deviation in Table \ref{tbl:trainingresults}. 
However, our main focus is on the test-set performances of the models that are presented in Table \ref{tbl:testresults}.

\spara{F1 Score:} This metric is the harmonic mean of the precision and recall.
The results of {\fone} are presented in Table \ref{tbl:testresults}.
We present the mean and standard deviation of the models' performances over 20 consecutive matches.

First, we compare the performances of the models when all features of Section \ref{sec:features} are used (rows 1-5) and when the best subset of features is used (rows 6-8).
We notice that there is improvement in the models' performances when the best features are used.
For instance in the case of the {\linear} and {\linearplus} the performance increases up to ${\scriptsize \sim}4\%$.
The conclusion of this observation is two-fold; (i) selecting the best subset of features boosts the models' performances, (ii) the feature engineering described in Section \ref{sec:features} and the features that we propose are overall very effective for the prediction of balanced matches.

Now, we focus on the individual comparisons between the different models.
Note that {\nnplus} achieves the best {\fone} performance (row 8) in both the 3v3 and 6v6 datasets.
An interesting observation is that even though {\nnplus} demonstrates the best performance during testing, its performance is not significantly higher than the performance presented by the much simpler {\linearplus} model (at most $4\%$ more). 
Overall, we see that the performances of {\nnplus}, {\linearplus} and {\rfplus} are close. 
Finally, for the 3v3 and 6v6 datasets we see that the {\fone} scores of {\dummy} and {\avg} are $0.00$ and $0.61$, respectively.
The {\dummy} model classifies all the matches as unbalanced, hence the zero {\fone} score. 
In both cases however, the conclusion is that simply using the average skill as a feature is not a good predictor of match balance.

\begin{table}
\centering
\footnotesize
  \begin{tabular}{|l|l|l|l|l|l|l||l|l|l|l|}
    \hline
      \multicolumn{1}{|c|}{\textbf{Model}} &
      \multicolumn{2}{c|}{\textbf{F1}}\\
    & 3v3 & 6v6\\
    \hline
    {\dummy}
    & $0.00\;(\pm 0.00)$ & $0.60\;(\pm 0.01)$ \\
    
    {\avg} 
    & $0.00\;(\pm 0.00)$ & $0.60\;(\pm 0.01)$\\
    
    {\linear}
    & $0.53\;(\pm 0.02)$ & $0.68\;(\pm 0.03)$ \\
    
    {\rf} 
    & $0.56\;(\pm 0.02)$ & $0.61\;(\pm 0.03)$\\
    
    {\nn} 
    & $0.59\;(\pm 0.02)$ & $0.68\;(\pm 0.02)$\\
    \hline
    
    {\linearplus} 
    & $0.60\;(\pm 0.02)$ & $0.68\;(\pm 0.02)$\\
    
    {\rfplus} 
    & $0.58\;(\pm 0.01)$ & $0.64\;(\pm 0.03)$\\
    
    {\nnplus}
    & $\bf 0.62\;(\pm 0.02)$ & $\bf 0.71\;(\pm 0.02)$ \\
    \hline
  \end{tabular}
  \caption{Test-set performance of models when predicting competitive balance. The results are averaged over 20 matches. 
		\label{tbl:testresults}}
\end{table}

\begin{table}
\centering
\footnotesize
  \begin{tabular}{|l|l|l|l|l|l|l|l|l||l|l|l|l|}
  \hline
    \multicolumn{1}{|c|}{\textbf{Model}} &
      \multicolumn{2}{c|}{\textbf{F1}}\\
    & 3v3 & 6v6\\
    \hline
    {\logistic}
    & $0.54\;(\pm 0.02)$ & $0.56\;(\pm 0.03)$\\
    
    {\soft}
    & $0.57\;(\pm 0.01)$ & $0.59\;(\pm 0.02)$\\
    
    \hline
    
    {\logplus}
    & $0.55\;(\pm 0.01)$ & $0.56\;(\pm 0.02)$\\
    
    {\softplus}
    & $0.55\;(\pm 0.01)$ & $0.62\;(\pm 0.02)$\\
    \hline
    
     {\delal}
    & $\bf 0.59\;(\pm 0.01)$ & $\bf 0.70\;(\pm 0.01)$\\
    
    \hline
  \end{tabular}
  \caption{Test-set performance of models when predicting probability of winning. The results are averaged over 20 matches.
		\label{tbl:classresults}}
\end{table}

\subsection{Why predict the score difference?}
\label{sec:esd}
The purpose of this section is first to demonstrate the effectiveness of using a competitive balance prediction model as opposed to a probability of winning prediction model.
The differences of the aforementioned models are presented in Section \ref{sec:predmodel}.

For this purpose, we define {\logistic}, {\logplus}, {\soft} and {\softplus} all of which are probability of winning prediction models and are trained to predict the probability that a team will win. 
{\logistic} and {\logplus} use the same features as {\linear} and {\linearplus}, respectively, but perform logistic regression, while {\soft} and {\softplus} use the same features and neural network architectures as {\nn} and {\nnplus}, respectively, but with an additional softmax layer that predicts the probability of winning.

Table \ref{tbl:classresults} demonstrates the performances of the probability of winning models on the test set.
Due to space limitations and given that the performance differences are pronounced we omit the corresponding results of the training set.
We compare the results of Table \ref{tbl:classresults} to the corresponding scores of Table \ref{tbl:testresults} where we consider the competitive balance prediction models. 
Specifically, we focus on the comparison of the following models; i) {\linear} with {\logistic}, ii) {\linearplus} with {\logplus}, iii) {\nn} with {\soft}, iv)  {\nnplus} with {\softplus}.
We see that using competitive balance models leads to higher {\fone} scores compared to predicting the probability of winning.
This is pronounced by the models' corresponding {\fone} scores which are overall much lower compared to Table \ref{tbl:classresults}.
The only exception is when comparing {\linear} with {\logistic} where the latter performs slightly better.

Furthermore, we perform a comparison of our proposed models with the probability of winning model proposed in \cite{delalleau2012beyond} denoted as {\delal} in row 5 of Table \ref{tbl:classresults}.
In that paper the authors present a neural network architecture with the following task; given two teams $A$ and $B$ predict the probability of team $A$ to win over team $B$.
Similar to the final score difference, we define a threshold $\omega$ to compute balanced and non-balanced matches from the probability of winning.
In particular, we consider the match to be balanced if $  | \textrm{Pr(team $A$ wins over team $B$)} - \frac{1}{2} |\leq \omega$, otherwise the match is not considered balanced.
Furthermore, since \cite{delalleau2012beyond} addresses a different game and the authors do not provide the exact feature and embedding descriptions, we use as input features to corresponding best subset of features as presented in Section \ref{sec:features}.
Overall when focusing on the {\fone} score of {\delal} in Table \ref{tbl:classresults} and comparing it to the corresponding score of {\nnplus} in Table \ref{tbl:testresults} we see that using the competitive balance models is more effective for the determination of balanced matches than using the probability of winning.
Another takeaway is that while {\nnplus} performs similar to {\delal}, we observe that the same applies to {\linearplus}, whereas this is not the case for {\logplus}.
We optimized this hyperparameter for best performance  of {\delal} model (obtained by $\omega = 0.3$).

\subsection{Training and inference times}
\begin{table*}
\centering
\footnotesize
  \begin{tabular}{|c|c|c|c|c|c|c|c|c|c|c|}
    \hline
    \multicolumn{1}{|c|}{\textbf{Prediction Model}} &
      \multicolumn{1}{|c|}{\textbf{Model}} &
      \multicolumn{2}{c|}{\textbf{Time 3v3}}\\
    & & Training & Inference\\
    \hline
    Competitive balance & {\dummy}
    & 1.0e-04 & 1.0e-05\;($\pm$ 0.0e-00)\\
    
    Competitive balance & {\avg}
    & 5.0e-02 & 5.0e-05\;($\pm$ 0.0e-00)\\
    
    Competitive balance & {\linear}
    & 8.2e+00 & 7.0e-05\;($\pm$ 0.0e-00)\\
    
    Competitive balance & {\rf} 
    & 9.9e+03 & 6.0e-03\;($\pm$ 7.0e-04) \\
    
    Competitive balance & {\nn}
    & 2.3e+02 & 2.1e-02\;($\pm$ 1.0e-02)\\
    
    Probability of winning & {\logistic}
    & 1.2e+02 & 6.0e-05\;($\pm$ 0.0e-00)\\
    
    Probability of winning & {\soft}
    & 3.7e+02 & 2.1e-02\;($\pm$ 5.0e-03)\\
    \hline
    
    Competitive balance & {\linearplus}
    & 5.6e+00 & 5.0e-05\;($\pm$ 0.0e-00)\\
    
    Competitive balance & {\rfplus} 
    & 8.3e+03 & 7.0e-03\;($\pm$ 1.0e-02)\\
    
    Competitive balance & {\nnplus}
    & 4.7e+02 & 2.4e-02\;($\pm$ 1.4e-02)\\
        
    Probability of winning & {\logplus}
    & 6.2e+01 & 6.0e-05\;($\pm$ 0.0e-00)\\
    
    Probability of winning & {\softplus}
    & 3.6e+02 & 2.3e-02\;($\pm$ 6.0e-03)\\
    
    \hline
    
    Probability of winning & {\delal}
    & 2.8e+01 & 2.3e-02\;($\pm$ 5.0e-03)\\
    
    \hline
  \end{tabular}
  \begin{tabular}{|c|c|c|c|c|c|c|c|c|c|c|}
    \hline
      \multicolumn{1}{|c|}{\textbf{Model}} &
      \multicolumn{2}{c|}{\textbf{Time 6v6}}\\
    & Training & Inference\\
    \hline
    {\dummy}
    & 1.0e-04 & 2.0e-05\;($\pm$ 0.0e-00)\\
    
    {\avg}
    & 3.5e-02 & 6.0e-05\;($\pm$ 0.0e-00)\\
    
    {\linear}
    & 5.2e-00 & 6.0e-05\;($\pm$ 0.0e-00)\\
    
    {\rf} 
    & 7.7e+03 & 8.0e-03\;($\pm$ 1.0e-03)\\
    
    {\nn}
    & 1.3e+02 & 2.4e-02\;($\pm$ 1.2e-02)\\
    
    {\logistic}
    & 8.7e+01 & 7.0e-05\;($\pm$ 0.0e-00)\\
    
    {\soft}
    & 1.2e+02 & 2.4e-02\;($\pm$ 7.0e-03)\\
    \hline
    
    {\linearplus}
    & 3.7e+00 & 8.0e-05\;($\pm$ 0.0e-00)\\
    
    {\rfplus} 
    & 6.2e+03 & 8.0e-03\;($\pm$ 1.0e-03)\\
    
    {\nnplus}
    & 1.6e+02 & 2.8e-02\;($\pm$ 8.0e-03)\\
        
    {\logplus}
    & 5.3e+01 & 7.0e-05\;($\pm$ 0.0e-00)\\
    
    {\softplus}
    & 2.4e+02 & 2.7e-02\;($\pm$ 1.0e-02)\\
    
    \hline
    
    {\delal}
    & 1.6e+02 & 2.7e-02\;($\pm$ 8.0e-03)\\
    
    \hline
  \end{tabular}
  \caption{Training time required for matches that occurred within a 3-month data period. Inference time averaged over 20 matches. The left table is for 3v3 matches and the right table is for 6v6 matches.
		\label{tbl:traininginfertimes}}
		\vspace{-.05in}
\end{table*}
Table~\ref{tbl:traininginfertimes} compares the training and inference times required by each of the prediction models.
Column 1 denotes the balance definition the corresponding model uses as described in Section \ref{sec:predmodel}.
The training time represents the time required to train each model over a series of matches that occurred within a 3-month period. 
The inference time of each model is the time required to make a prediction, averaged over 20 matches.
We report both the mean and the standard deviation of the models' running times in seconds.

Observe that the training and the inference times of the linear models are approximately $10$x and $100$x faster, respectively, compared to the corresponding times of the neural network-based models.
In online gaming taking the training and inference times into account is essential to provide high-quality service to the user without latency.
Therefore, even though the {\linearplus} model's {\fone} score is slightly lower than the one of the best-performance {\nnplus} model, in practice trading-off performance for speed can be essential for online gaming.
Note that the training and inference times of the 3v3 dataset are higher than the corresponding ones of the 6v6 dataset.
This is because the total number of matches in the 6v6 dataset are fewer compared to the 3v3 dataset.

\subsection{Significant features}

\begin{table*}
\centering
\footnotesize
\begin{tabular}{lcclll}
    \hline
    \textbf{$M$ Features} & \textbf{Coeff.} & \textbf{Coeff.}
     & \textbf{Description in Team Sports games} 
     \\& \textbf{3v3} & \textbf{6v6} & & & \\ \hline
        avg\_freq\_dropout   
    & $+1.128$ & $+1.164$ &
    Average dropout rate of the players in Teams 1 \& 2\\
    \hline 
    avg\_assists\_abs\_diff  
    & $+0.889$ & $+0.741$ & Absolute difference of the average number of assists between Teams 1 \& 2 \\
    \hline 
    avg\_freq\_defense & $+0.850$  
     &  $+0.918$ & Average rate of playing  defense among players of Teams 1 \& 2\\
    \hline 
    avg\_freq\_left & $+0.856$ 
     & $+0.504$ &
     Average rate of playing left wing among players of Teams 1 \& 2\\
     \hline 
    avg\_freq\_right & $+0.843$ 
     & $+0.494$ &
     Average rate of playing right wing among players of Teams 1 \& 2  \\
     \hline 
    avg\_freq\_wins 
    & $-0.334$ & $-0.178$ &
    Average win rate among players of Teams 1 \& 2 \\
    \hline 
    cnt\_players& $+0.117$ & $+0.142$ & 
    Number of human players in Teams 1 \& 2 in the  beginning of the match\\
    \hline
  \end{tabular}
    \caption{Statistical analysis of indicative most significant features. The fourth column describes these features for the online team sports online games used in our experiments.
		\label{tbl:significantfeatures}}
		\vspace{-.15in}
\end{table*}

\label{sec:sigfeatures}
This section provides a discussion on the important features of a match for predicting competitive balance.
Table \ref{tbl:significantfeatures} presents the most statistically significant features with their coefficient residuals and a corresponding brief description of their meaning in the team sports game.
For all features $p < 0.001$.

An interesting observation is that the frequency of dropouts (row 1), a common phenomenon in team online games, is a strong indicator of competitive balance.
A player dropping a game before it finishes results in having a team with less human players and therefore gives the lead to the opposing team.
As expected, statistics on the past actions of the players (row 2) are also significant for competitive balance, and in this case this action was particularly the assists that occurred in the game.
Rows 3-5 correspond to the role experience players have.
In the team sports game we are considering, the most important roles are defense, and right and left offense.
However, there are other roles in the game that appear to not be critical to the final outcome. 
In row 6 we observe a negative coefficient for the frequency of winning feature.
This implies that the largest the frequency of previous wins for a team the more unlikely it is that the match will end with a balanced score.
Finally, as expected, the number of players in each team at the beginning of the match (row 7) also seems to impact balance.
Potentially, this is because teams with less players are assigned with bots whose playing behavior significantly deviates from a human's, and thus can be more unexpected.

Note that while Table \ref{tbl:significantfeatures} presents the features with the largest coefficients in magnitude, we considered other features as well that had much smaller impact.
For instance, in addition to using the average of the features we also considered their corresponding standard deviation.
Furthermore, we also evaluated the skill ratings of the opposing teams.
The results showed that even though skill rating was not among the most significant features, we cannot draw conclusive results about its importance because the datasets we used comprise real matches between teams of close team skill ratings.
That said, we remark that performing a statistical significance test after the proposed model has been deployed in the matchmaking system, could provide us with potentially deeper insights, even though we expect the presented results to mostly hold.

In Table \ref{tbl:significantfeatures}, column 4 provides the descriptions of some of the most significant features of the team sports online game that we are investigating.
In addition to the statistical analysis of Table \ref{tbl:significantfeatures} we created correlation matrices to identify high amount of correlations that would suggest unreliable prediction estimates and removed these features from the dataset, which are ommitted however due to lack of space.
Finally, we used the \emph{recursive feature elimination} feature selection method and the results of the statistical analysis to decide the best subset of features for each of the baseline methods (when applicable) and for our proposed model presented in Section \ref{sec:baseline}.
An interesting observation was the common consensus between the feature selection methods and the different models about the significant features for competitive balance.
\vspace{-0.5pt}
\section{Concluding Remarks}
In this paper, we have examined the problem of making matches that exhibit competitive balance. Through simulations on an online team sports game published by Electronic Arts (EA), we demonstrated that regressing the final score difference on carefully designed player, team and match features followed by a binary threshold can significantly outperform aggregate skill-based models in predicting match balance. Our approach provides insight into how simple, generalizable attributes of players and teams can be used to better capture in-game dynamics that impact match outcomes. 
We also show that using a linear model with the specific features can lead to computational savings ranging by orders of magnitude with a small sacrifice in predicting performance.

A main focus of this work is to provide insight to game designers on how to improve the quality of online team games through better matchmaking.
We believe that the presented definitions, features, prediction models and experiments can be utilized by game designers for predicting competitive balance in other types of online games.
We have seen how the proposed models generalize from 3v3 to 6v6 games, but it still remains to see how it can be evaluated in different sports games and larger teams.
Finally, note that the models and experiments illustrated in this paper are based on existing player data. 
The approaches described here have not yet been deployed in a live matchmaking service.
However, they provide the hypotheses for A/B testing once they are deployed in the game.

\end{document}